\documentstyle[]{ptptex}

   \def\lesssim{\mathrel{\hbox{\rlap{\hbox{\lower4pt\hbox{$\sim$}}}\hbox{$<$}}}}
   \def\gtrsim{\mathrel{\hbox{\rlap{\hbox{\lower4pt\hbox{$\sim$}}}\hbox{$>$}}}}
   \newcommand{\bm}[1]{\mbox{\boldmath$#1$}}
   \newcommand{\mal}[1]{\stackrel{_\circ}{#1}}
   \newcommand{\kaco}[1]{\left\langle{#1}\right\rangle}

\title{\bf On a Growing Transverse  Mode as a Post-Newtonian 
Effect in the Large-Scale Structure Formation}

\author{Masahiro Takada
\footnote{E-mail address: takada@astr.tohoku.ac.jp}  
and   Toshifumi Futamase
\footnote{E-mail address: tof@astr.tohoku.ac.jp}
}

\inst{Department of Astronomy, Faculty of Science,
Tohoku University, Sendai, Miyagi  980-8578}
\abst{
 We point out the existence of a new type of growing transverse mode 
in the gravitational instability. This appears as a post-Newtonian effect 
to Newtonian dynamics. 
We demonstrate this existence 
 by formulating the Lagrangian perturbation theory 
in the framework of the cosmological 
post-Newtonian approximation in general relativity. 
Such post-Newtonian order effects might produce characteristic 
appearances of large-scale structure formation, for example, 
through the observation of anisotropy of the cosmic microwave 
background  radiation (CMB).  
}
\begin{document}

\maketitle

\section{Introduction}

The investigation of nonlinear, large-scale structure formation of
the universe has been a major research subject in cosmology. 
As long as the scales of the density fluctuations 
are much smaller than the horizon scale, the Newtonian approximation 
is sufficiently  accurate to describe their evolution up to the 
non-linear 
regime, where the density contrast  becomes larger than unity.
On the other hand, the region in which the large-scale 
structure is observed 
is steadily growing. For example, SDSS (Sloan Digital Sky
Survey) will cover over a region of several hundred megaparsecs.
It is not clear  if the application of  Newtonian 
theory is appropriate for  such a wide region of spacetime.
In fact  fluctuations relevant for such large-scale structures 
are not always  much smaller than horizon scales in the past. 
For example, the horizon scale at the decoupling
time is on the order of $ct_{\mbox{{\scriptsize dec}}}
(1+z_{\mbox{{\scriptsize dec}}})\sim
80h^{-1}\mbox{Mpc}$ in the present physical length. 
This suggests that we must
 employ a relativistic description 
for the evolution of fluctuations larger than or 
equivalent to such scales.

Thus to understand the evolution of the large-scale structure of the
universe, it is important and necessary to have
some formalism to clearly evaluate the effect of general
relativistic corrections to the  Newtonian dynamics. 
Naturally, we also require the formalism to agree with  the gauge invariant
linear theory developed by Bardeen\cite{bardeen},and 
 Kodama and 
Sasaki\cite{ks} in the linear regime.  

We develop such a formalism based on the Lagrangian perturbation 
theory in the post-Newtonian (PN) framework in which 
it has been shown that 
the formalism can be applied even for  perturbations 
larger than the present horizon scale $\sim 2000h^{-1}\mbox{Mpc}$.\cite{tf}
The Newtonian Lagrangian picture has also been developed 
by Buchert,\cite{buchert89,buchert92,buchert94} where it is expected that the 
approach gives a good approximation up to a certain stage of the 
non-linear 
regime.\cite{buchertW,mbw,weiss}  
There have also been some studies based on 
 relativistic Lagrangian
perturbation theory.\cite{kasai95,mt} 
However, because of the gauge condition adopted in these studies, 
namely,  the synchronous comoving coordinates, it is not easy to 
have contact with the Newtonian Lagrangian approach fully developed 
and used for the numerical simulation. 
Thus we shall study the Lagrangian perturbation 
in 
 coordinates where  comparison with the  Newtonian case is 
most easily carried out.  

\section{Cosmological post-Newtonian approximation}

As stated above, we wish to develop the Lagrangian 
perturbation theory 
formulated in a coordinate system which has a straightforward 
Newtonian limit. 
For this purpose it is convenient to use the (3+1) formalism motivated 
by  the following considerations. 
Let us first assume that there exists a congruence of time-like 
worldlines from which the spacetime looks isotropic. We shall call the 
family of the worldlines basic observers, who see no dipole component 
 of the 
cosmic microwave background radiation (CMB). We can regard that 
any one of these observers at the spacetime point $x$ moves with
4-velocity $n^\mu(x)$, without loss of generality. The tangent vector 
$n^\mu$ is normalized as $n_\mu n^\mu=-1$. These observers are used to 
foliate the spacetime by their simultaneous surfaces: $t=\mbox{const}$.
These considerations imply  that  matter, such as a galaxy, 
moves with a velocity in our coordinates 
according to  the geodesic equation.
Hence we naturally expect that  the dynamics of 
 nonrelativistic matter  is described by the
cosmological Newtonian picture at the lowest order. 
Fortunately, there have been some studies on the post-Newtonian 
approximation applied to cosmology\cite{futamase96,sa} in the (3+1)
formalism. 
We shall follow 
these approaches to work in the following coordinates:
\begin{equation}
ds^2 = -(\alpha^2-\beta_{i} \beta^{i})(cdt)^2+2\beta_{i}(cdt)dx^{i}
+ a^2(t) [ (1-2 \psi )\tilde{\gamma}^{(B)}_{ij}+h_{ij} ]dx^{i}dx^{j},
\label{eqn:aa}
\end{equation}
where $\tilde{\gamma}^{(B)ij}h_{ij}=0$.  We employ 
 gauge conditions termed the ``longitudinal gauge'' or 
``Newtonian gauge'' with regard to 
the scalar modes of the metric perturbations:\cite{bert}     
\begin{eqnarray} 
\tilde{\gamma}^{(B)ij}\tilde{D}^{(B)}_{j}\beta_i=0, \hspace{2em}
\tilde{\gamma}^{(B)jk}\tilde{D}^{(B)}_{k}h_{ij}=0.\label{eqn:a}
\end{eqnarray}
Here $\tilde{D}^{(B)}_i$ denotes the 
covariant derivative with respect
to the background metric $\tilde{\gamma}^{(B)}_{ij}$, for
 which 
we take the spatial metric of the Friedmann-Robertson-Walker
(FRW) geometry as $\tilde{\gamma}^{(B)}_{ij}
=\delta_{ij}/( 1+{\cal K}r^2/4)^2$. 
The quantity 
 ${\cal K}$ is the curvature parameter of FRW models and 
$r^2=x_1^2+x_2^2+x_3^2$. 
We treat the metric perturbations as  small quantities, 
so we are able to regard tensorial quantities in
our equations as tensors with respect to the background spatial metric
$\tilde{\gamma}^{(B)}_{ij}$.
The above  condition ({\ref{eqn:a}}) guarantees that 
$h_{ij}$ and $\beta_i$  contain only the tensor mode,  
which represents the freedom of the
gravitational wave in the PN order, 
 and the vector mode, respectively.

The explicit forms of Einstein equations in terms of the 
above variables 
can be found in previous works.\cite{futamase96,sa}
In the following,  we only consider the Einstein-de Sitter background 
 universe (${\cal K}={\Lambda}=0$) 
and use Cartesian coordinates for the sake of simplicity: 
\begin{equation}
H(t)^2\equiv\left( \frac{\dot{a}(t)}{a(t)}\right)^2=\frac{8 \pi G \rho_b(t)}{3},
\label{eqn:at}
\end{equation}
where $H(t)$ and $\rho_b(t)$ are the Hubble parameter and 
the homogeneous density of the background FRW universe, respectively. 
For simplicity, we assume that 
the above equation is derived by the usual averaging
method\cite{futamase96}  and 
we may regard $\rho_b$ as an average value over 
the volume as large as the horizon scale $(ct)^3$:
$\kaco{\rho}_{(ct)^3}\equiv\rho_b(t)$.

The metric required for the usual Eulerian Newtonian
picture in the perturbed FRW universe 
is known to take the following form:
\begin{equation}
ds^2=-\left(1+\frac{2 \phi_N}{c^2}\right)c^2dt^2
+a^2(t)\left(1-\frac{2 \phi_N}{c^2}\right)
\delta_{ij}dx^idx^j.
\label{eqn:c}
\end{equation}
The quantity  $\phi_N $ is a cosmological Newton-like  gravitational potential 
related to the matter density 
fluctuation field $\delta_N$ via the Poisson equation
\begin{equation}
\Delta_x \phi_N=4\pi G a^2\rho_b\delta_N,
\end{equation} 
where 
$\delta_N({\mbox{\boldmath $x$}},t)\equiv 
(\rho_N({\mbox{\boldmath $x$}},t)-\rho_b(t))/
\rho_b(t)$, and $\Delta_x$ is the Laplacian.
The above metric is usually used to give an accurate description of 
the trajectories
of nonrelativistic fluid elements on scales
much smaller than the Hubble distance $cH^{-1}$.

We expand the basic equations in powers of $c^{-1}$ 
to obtain  the post-Newtonian approximation with the condition that 
the lowest order metric takes the above Newtonian form 
(\ref{eqn:c}). Thus the PN terms of all metric variables used 
in this paper may be expanded as follows:\cite{ch} 
\begin{eqnarray}
&&\alpha = 1+\frac{\phi}{c^2}+\frac{\alpha^{(4)}}{c^4}+O(c^{-6}), \nonumber \\
&&\psi= \frac{\psi^{(2)}}{c^2}+\frac{\psi^{(4)}}{c^4}+O(c^{-6}),  \nonumber \\
&&\beta^i=\frac{\beta^{(3)i}}{c^3}+O(c^{-5}), \nonumber \\
&& h_{ij}=\frac{h^{(4)}_{ij}}{c^4}+O(c^{-5}), \label{eqn:bb}
\end{eqnarray}
where $\phi$ is the Newtonian  gravitational potential as stated below.    
It should be noted that we consider only the PN expansion of $\beta^i$,
not of $\beta_i$. 
If we adopt a dust model $T_{\mu\nu}=\rho
c^2u_\mu u_\nu$ for the energy-momentum tensor of matter,  
  the metric up to order  $c^{-2}$ 
agrees with (\ref{eqn:c}) by substituting the above expressions 
into Einstein equations.
From the other substitutions,  
we can derive the relations between the metric perturbations and
the matter variables in each order of $c^{-n}$. 
We shall here present only relevant equations in our calculations.
The lowest order field equation  gives us
\begin{eqnarray}
\Delta_x\phi=\Delta_x\psi^{(2)}=4 \pi G a^2(\rho -\rho_b). \label{eqn:bf}
\end{eqnarray}
The field equations at the next order, that is, at  the first PN order 
take the following forms for $\alpha^{(4)}$ and $\beta^{(3)i} $: 
\begin{eqnarray}
&&\Delta_x\alpha^{(4)}=\phi,_k\phi,_k+4\pi G a^2 \left( 2\rho a^2 v^2 -\rho \phi +3 \rho_b 
\phi \right)-3a^2\left(\frac{\partial^2 \phi}{\partial
t^2}+3\frac{\dot{a}}{a}\frac{\partial \phi}{\partial t}\right),
\label{eqn:bk} \\
&&\Delta_x\beta^{(3)i}
=16 \pi G a^2 \rho v^i+4\left(\frac{ \partial \phi,_i}{\partial t}+
\frac {\dot{a}}{a}\phi,_i\right) ,
\label{eqn:bm}
\end{eqnarray}
where we have used the Newtonian order equations (\ref{eqn:bf}) and 
the gauge
conditions (\ref{eqn:a}), and
$v^i$ is defined as $v^i/c=u^i/u^0$. We 
  note that  $v^i$ represents a peculiar velocity of a fluid element 
in the comoving frame, so the physical peculiar velocity is $av^i$.

Similarly, the material equations up to the first PN order 
become
\begin{eqnarray}
&&\frac{\partial }{\partial t}\left[ \rho a^3 \left\{ 1
+\frac{1}{c^2}\left( \frac{1}{2}a^2v^2
-3\phi \right) \right\} \right]\!+\!\frac{\partial }{\partial x^i}\!\left[
\rho a^3 v^i \left\{ 1+\frac{1}{c^2}\left(\frac{1}{2}a^2v^2-3\phi\right) 
\right\}\right] \nonumber \\
&&\hspace{24em}\!+O(c^{-4})\! =0, \label{eqn:bt} \\
&&\frac{\partial v^i}{\partial t}+v^j
\frac{\partial v^i}{\partial x^j}+2\frac{\dot{a}}{a}v^i
+\frac{1}{c^2}v^i\left[ \frac{\partial }{\partial t}\left(\frac{1}{2}a^2v^2-3\phi 
\right)+v^j\frac{\partial}{\partial x^j}\left(\frac{1}{2}a^2v^2
-3\phi\right)
\right] \nonumber \\
&&\hspace{5em}=-\frac{1}{a^2}\phi,_i-\frac{1}{c^2}\frac{1}{a^2}\left[ \alpha^{(4)},_i+
3\phi \phi,_i+a^2 v^2\phi,_i
+\frac{\partial }{\partial t}(a^2 \beta^{(3)i})\right.\nonumber\\ 
&&\hspace{16em}\left.+a^2v^j(\beta^{(3)i},_j-\beta^{(3)j},_i)\frac{{}}{{}}\!
 \right] +O(c^{-4}). \label{eqn:g}
\end{eqnarray}
The lowest order in these two equations reduces to
the Newtonian Eulerian equations of 
hydrodynamics for a pressureless fluid, 
and the terms of order $c^{-2}$ provide the first PN corrections. 
Equations (\ref{eqn:bt}) and (\ref{eqn:g}) are our basic equations for
the Lagrangian approach to the trajectory field of  matter fluid elements.

\section{PN Lagrangian perturbation approach to  large-scale
structure formation in the perturbed FRW universe}

\subsection{Basic equations for the transverse part of the trajectory
field of the fluid elements}

We shall rewrite the above set of post-Newtonian equations in the Lagrangian 
coordinates and then solve them perturbatively 
by extending the formalism in the Newtonian theory
developed by Buchert.\cite{buchertpre} Namely we concentrate on the integral 
curves $\bm{x}=\bm{f}(\bm{X},t)$
of the velocity field $\bm{v}(\bm{x},t)$:
\begin{equation}
\frac{d \bm{f}}{dt}\left(=\dot{\bm{f}}\right) \equiv \left.\frac{\partial \bm{f}}
{\partial t}\right|_X
=\bm{v}(\bm{f},t) ,\hspace{1.5cm} \bm{f}(\bm{X},t_I)\equiv \bm{X},
\label{eqn:map}
\end{equation}
where $\bm{X}$ denotes the Lagrangian coordinates 
which label each fluid element, \bm{x} is 
the position of such an element in Eulerian space at time $t$, 
and $t_I$ is the initial time when the  Lagrangian coordinates 
are defined. 
It should be noted that we introduce the Lagrangian coordinates on the
comoving coordinate because we have already derived basic equations by
using  the FRW metric defined with 
the comoving coordinates. 
As long as the mapping $\bm{f}$ is invertible,
 we can give the inverse of the deformation tensor $f_{i|j}$, which
is written in terms of  variables $(\bm{X},t)$:
\begin{equation}
\left.\frac{\partial X^i}{\partial x^j}\right|_t \equiv h_{i,j}(\bm{X},t) 
=\frac {1}{2J}\epsilon_{iab}\epsilon_{jcd}f_{c|a}f_{d|b} ,\label{eqn:dm}
\end{equation}
where $J$ is the determinant of the deformation tensor  $f_{i|j}$,
and  $\epsilon_{ijk}$ is a totally antisymmetric quantity with
$\epsilon_{123}=+1$. 
The comma and the vertical slash  in the subscript denote 
partial differentiation with respect 
to the Eulerian coordinates and the Lagrangian coordinates, 
respectively. 

The continuity equation (\ref{eqn:bt}) then 
may be rewritten as 
\begin{equation}
\frac{d}{dt}\left[ \rho a^3\left(1+ \frac{1}{c^2}A(\bm{X},t)
\right)J(\bm{X},t)\right]
+O(c^{-4})=0 ,\label{eqn:da}
\end{equation}
where 
\begin{equation}
A(\bm{X},t)\equiv \frac{1}{2}a^2(t)\dot{\bm{f}}^2
-3\phi(\bm{X},t).  
\end{equation}
Thus the density field is integrated exactly up to the first PN order 
in the Lagrangian picture as in the Newtonian case:
\begin{equation}
\rho(\bm{X},t)a^3J(\bm{X},t)=\left(1+ \frac{1}{c^2}A(\bm{X},t)\right)^{-1}
\mal{\rho}(\bm{X})\mal{a}^3\left(1+\frac{1}{c^2}\mal{A}(\bm{X})\right)
+O(c^{-4}),
\label{eqn:db}
\end{equation} 
where the quantities with $\mal{}$, such as $\mal{a}$,
 are the quantities 
at the initial time $t_I$,  
$\Delta_X\!\!\mal{\phi}=4\pi G\! \mal{a}^2\!(\mal{\rho}-\mal{\rho_b})$ 
with $\Delta_X$ representing the Laplacian with respect to the Lagrangian 
coordinate, and  $\mal{J}=1$.

The trajectory field  \bm{f} is determined by solving the PN equations 
of motion (\ref{eqn:g}). 
As in the Newtonian case,\cite{buchertpre} 
we consider the rotation and the divergence  of the equations 
of motion  with respect to the Eulerian coordinates,
 respectively. 
In the Lagrangian picture, the former set becomes basically  
the evolution equation of the transverse part of the trajectory 
field, and the latter becomes the evolution equation 
 of its longitudinal part. 
We shall here concentrate on the transverse part at the PN order. 
The longitudinal  part will be discussed elsewhere in detail.\cite{tf}

So by operating with rotation on Eq. (\ref{eqn:g}) with respect to 
the Eulerian coordinate we obtain
\begin{eqnarray}
&&\hspace{0em}\epsilon_{ijk}\frac{\partial }{\partial x^j}\!\left[
\ddot{f}^k+2\frac{\dot{a}}{a}\dot{f}^k+\frac{1}{c^2}
\dot{f}^k\dot{A}\right]=-\frac{1}{c^2}\frac{1}{a^2}
\epsilon_{ijk}\frac{\partial }{\partial x^j}\left[ 
a^2v^2\phi_{,k}\right. \nonumber \\
&&\hspace{8em}\left.+\frac{\partial }{\partial t}\!\left(a^2\beta^{(3)k}
\right)+a^2v^l\left(\beta^{(3)k}{}_{,l}-\beta^{(3)l}{}_{,k}\right)
\right]\!+O\!\left(c^{-4}\right)\!,\label{eqn:pnvort}
\end{eqnarray}
According to the usual procedure in the Lagrangian formalism, 
 we can modify the above equation by using
Eqs. (\ref{eqn:map}) and (\ref{eqn:dm}) in the following form: 
\begin{eqnarray}
\lefteqn{\epsilon_{abc}f_{j|a}f_{i|b}\left[\ddot{f}_{j|c}+
2\frac{\dot{a}}{a}\dot{f}_{j|c}\right]=
-\frac{J}{c^2}\left[\frac{1}{a^2}\frac{\partial }{\partial t}
\!\left(a^2\epsilon_{ijk}\beta^{(3)k}{}_{,j}\right)
\right.}\nonumber \\
&&\hspace{12em}\left.+\epsilon_{ijk}\frac{\partial }{\partial x^j}\!
\left\{v^l\left(\beta^{(3)k}{}_{,l}-\beta^{(3)l}{}_{,k}\right)\right\}
+\frac{2}{J}\epsilon_{abc}f_{i|b}\dot{f}_l
\dot{f}_{l|c}\phi_{|a}\right]\nonumber \\
&&\hspace{12em}-\frac{1}{c^2}\epsilon_{abc}f_{j|a}
f_{i|b}\left(\dot{f}_{j|c}\dot{A}+\dot{f}_j
\dot{A}_{|c}\right)+O\!\left(c^{-4}\right), \label{eqn:transtrajdiff}
\end{eqnarray}
where we have not transformed  the terms including the shift vector
$\beta^{(3)i}$  in Eq. (\ref{eqn:transtrajdiff}) 
into their forms in the  Lagrangian picture,
because the expression in the Eulerian picture 
will make it easier to understand the expansion of 
the above equation in later discussion. 
We note that the right-hand side of Eq. (\ref{eqn:transtrajdiff}) 
is of order $c^{-2}$; namely, it has  only 
quantities of  first PN order. 

The structure of the above equation allows us to solve for  
the displacement vector iteratively up to any desired 
order in $c^{-n}$ in terms of the lower order displacement 
vectors and metric variables.

\subsection{The solution of the transverse part at PN order}

The  solution of Eq. (\ref{eqn:at})
in an Einstein-de Sitter background can be chosen as
\begin{equation}
a(t)=\left(\frac{t}{t_0}\right)^{2/3},
\end{equation}
where $t_0$ denotes the present time.
We use this normalization for the scale factor since it 
makes the physical interpretation of the solution more 
transparent, partly because 
the length in the comoving frame is equal to the physical 
length at the present
universe with this choice. Furthermore, 
we shall assume that the initial hypersurface exists at 
 sufficiently early time 
so that the initial density contrast field is much smaller than unity. 
This allows us to adopt the order of the initial density contrast, say  
$\lambda$, as a 
new perturbation parameter.

The parameter $\lambda$ is assumed to be small and dimensionless. 
We formally split the initial density field accordingly:
\begin{equation}
\mal{\rho}\!(\bm{X})=\mal{\rho}_b+\lambda
 \mal{\rho}_b\,\mal{\delta}\!(\bm{X}),
 \hspace{2em}\kaco{\mal{\rho}\!(\bm{X})}_{\mal{V}}=\mal{\rho}_b,
\end{equation}
where $\mal{\delta}$ denotes the initial density contrast field. 
This choice of the initial data is adequate, because the density
field at some given time need not be perturbed in the Lagrangian framework.

We assume that the vorticity of the Newtonian displacement
vector is negligible, which seems to be
reasonable.\cite{buchert92,buchert94,peebles}
We consider the following ansatz for the transverse perturbations 
at the first PN order superposed on the solved Newtonian irrotational 
trajectory field:
\begin{eqnarray}
&&\hspace{-2em}f^i(\bm{X},t)={}^{(N)}\!f^i(\bm{X},t)+\!\frac{1}{c^2}
\!\left(\lambda Q_z^T(t){}^{(PN)}\!\Xi^{(1)}_i(\bm{X})+\lambda^2
Q_{zz}^T(t){}^{(PN)}\!\Xi^{(2)}_i\left(\bm{X},t\right)
\right),\nonumber \\ \label{eqn:transtraj}
\end{eqnarray}
where 
\begin{subequations}\label{eqn:newtrajsol}
\begin{equation}
{}^{(N)}\!f^i(\bm{X},t)=X^i+\lambda q_z(t)\Psi^{(1)}_{N|i}(\bm{X})
+\lambda^2q_{zz}(t)\Psi^{(2)}_{N|i},\label{eqn:newtrajsol1}
\end{equation}
with
\begin{eqnarray}
&&q_z=\left(\frac{3}{2}\right)
\left[\left(\frac{t}{t_I}\right)^{2/3}-1\right],\\
&&q_{zz}=\left(\frac{3}{2}\right)^2\left[-\frac{3}{14}\left(\frac{t}{t_I}\right)^{4/3}
+\frac{3}{5}\left(\frac{t}{t_I}\right)^{2/3}-\frac{1}{2}+\frac{4}{35}\left(
\frac{t}{t_I}\right)^{-1}\right],
\end{eqnarray}
and
\begin{eqnarray}
&&\Delta_X\Psi^{(1)}_N=-\frac{2}{3}\mal{\delta}_g,\label{eqn:newvelpot}\\
&&\Delta_X\Psi^{(2)}_N=\Psi^{(1)}_{N|ii}\Psi^{(1)}_{N|jj}
-\Psi^{(1)}_{N|ij}\Psi^{(1)}_{N|ij}\label{eqn:newvelsecond}. 
\end{eqnarray}
\end{subequations}
The initial conditions are imposed on the time coefficients $Q^T$
 by the definition of the Lagrangian
coordinate (\ref{eqn:map}): $Q^T_z(t_I)=Q^T_{zz}(t_I)=0$. 
The quantity $\mal{\delta}_{g}$ is defined as $\mal{\delta}-2\mal{\phi}/c^2$   
and denotes the initial gauge invariant density
contrast field.\cite{sa,tf} 
${}^{(N)}\!f^i$ is the Newtonian solution up to the  second order in
$\lambda$,\cite{buchert94} and 
we have here ignored the PN part of the longitudinal part 
(discussed elsewhere\cite{tf}).
The PN transverse displacement vector at each order is constrained by
\begin{equation}
\nabla_X\cdot {}^{(PN)}\!\bm{\Xi}^{(1)}=\nabla_X\cdot
 {}^{(PN)}\!\bm{\Xi}^{(2)}=0.
\end{equation}
Before considering the PN rotational flow in our coordinates, 
we remark that  
we can express the Newtonian gravitational potential
constrained by Eq. (\ref{eqn:bf}) in terms of the above Newtonian
solution  ${}^{(N)}\!f^i$\cite{tf} following the Lagrangian perturbation 
formalism: 
\begin{eqnarray}
&&\phi_N(\bm{X},t)=-\lambda \frac{1}{t_I^2(1+z_I)^2}\Psi_N^{(1)}
-\lambda^2\left[ \left( 2a \dot{a} \dot{q}_{zz} 
+a^2\ddot{q}_{zz}\right)
\Psi_N^{(2)}\right.\nonumber \\
&&\hspace{14em}\left.+q_z\left( 2a \dot{a} \dot{q}_{z} 
+a^2\ddot{q}_{z}\right)\tilde{\phi}
(\bm{X})\right] +O(\lambda^3),\label{eqn:newtonpot}
\end{eqnarray}
where 
\begin{equation}
\tilde{\phi}(\bm{X})\equiv- \int \!d\bm{Y}\frac{
[\Psi^{(1)}_{N|ij}(\bm{Y})
\Psi^{(1)}_{N|i}(\bm{Y})]_{|j}}{4\pi |\bm{X}-\bm{Y}|}.
\end{equation}
%
In the above derivation, we have used the Green function
$-1/(4\pi |\bm{X}-\bm{Y}|)$  of the Laplacian
motivated by the following considerations. 
As seen in Eq.(\ref{eqn:bf}), the Newton-like gravitational potential 
in the cosmological situation 
is generated by the density contrast $\delta$. Then  
 we can safely  express such quantities  using the Green function
of the Laplacian, because the integrals over 
the horizon scale converge to a definite value by assuming that 
the density contrast field obeys 
periodic boundary condition on the horizon scale. 
We have found that the cosmological PN treatment to the large-scale
structure formation is valid even for perturbations larger than 
the present horizon scale.\cite{tf} 

We need  Eq. (\ref{eqn:transtrajdiff}) to solve the transverse part
$Q^T\Xi_i$  
of the
PN displacement vector. 
Naturally, the order $c^0$ in this equation
produces Newtonian  counterparts 
in the case that the trajectory field is introduced on the comoving
coordinates.\cite{buchert92,buchert93,buchert94} 
Then, if we assume $J_N > \tilde{J}_{PN}/c^2$, where 
 $\tilde{J}_{PN}$ is a term of  order $c^{-2}$  in expanding
$J$,  and  take into account the  order of $\lambda$ in each 
term on the right-hand side,
Eq. (\ref{eqn:transtrajdiff}) becomes  
\begin{eqnarray}
&&\epsilon_{abc}f_{j|a}f_{i|b}\left[ \ddot{f}_{j|c}+
2\frac{\dot{a}}{a}\dot{f}_{j|c}\right]
 =-\frac{J_N}{c^2}\frac{1}{a^2}\frac{\partial }{\partial t}
\!\left(a^2\epsilon_{ijk}\beta^{(3)k}{}_{,j}\right)\nonumber \\
&&\hspace{8em}-\frac{J_N}{c^2}\epsilon_{ijk}\frac{\partial}{\partial x^j}\!
  \left[{}^{(N)}\!\dot{f}^l(\beta^{(3)k},_l-\beta^{(3)l},_k)\right]\!\!
  +O\!\left(c^{-4},\lambda^3\right), \label{eqn:transeqns}
\end{eqnarray}
where 
we have neglected  terms like $\phi_N v^2_N$ in our approximation 
because they are of  order $\lambda^3$ at most, and 
$J_N\equiv \mbox{{\rm det}}({}^{(N)}\!f_{i|j})$.
This equation shows that  the 
shift vector constrained by the gauge condition (\ref{eqn:a})  
generates the first PN transverse flow. 
Inserting the ansatz (\ref{eqn:transtraj}) into
Eq. (\ref{eqn:transeqns}),  we obtain the following
differential equation for solving the
transverse velocity potential  of the first PN displacement vector
at the order $c^{-2}$:
\begin{eqnarray}
&&\lambda
\left(\ddot{Q}^T_z 
+2\frac{\dot{a}}{a}\dot{Q}^T_z\right)\left(\nabla_X\times
{}^{(PN)}\!\bm{\Xi}^{(1)}\right)_i 
 +\lambda^2\left(\ddot{Q}^T_{zz}+2\frac{\dot{a}}{a}\dot{Q}_z
\right)\left(\nabla_X\times{}^{(PN)}\!\bm{\Xi}^{(2)}\right)_i\nonumber \\
&&\hspace{2em}=
-\frac{J_N}{a^2}\frac{\partial }{\partial t}\left(a^2
  \left(\nabla_X\times\bm{\beta}^{(3)}\right)_i\right)
 -J_N\epsilon_{ijk}\frac{\partial}{\partial x^j}\!
  \left[v_N^l(\beta^{(3)k},_l-\beta^{(3)l},_k)\right]\nonumber \\
&&\hspace{3em}  -\lambda^2\left(\ddot{q}_z+2\frac{\dot{a}}{a} 
\dot{q}_{z}\right)\epsilon_{ijk}\Psi^{(1)}_{N|jl}Q^T_z{}^{(PN)}\!\Xi^{(1)}_{l|k}
\nonumber    \\
&&\hspace{3em}-\lambda^2 q_z\left(\epsilon_{ijk}\Psi^{(1)}_{N|kl}
+\epsilon_{jkl}\Psi^{(1)}_{N|ik}\right)\left(
\ddot{Q}^T_z+2\frac{\dot{a}}{a}\dot{Q}^T_z
\right){}^{(PN)}\!\Xi^{(1)}_{l|j}
  +O\!\left(\lambda^3\right). \label{eqn:pntransdiff}
\end{eqnarray}
The form of this equation shows that it  can be solved 
if we  express the shift vector $\beta^{(3)i}$ in terms of
already-known Newtonian 
trajectory field ${}^{(N)}\!f^i$.

Thus we first must  solve the  shift vector with respect to
the Lagrangian coordinate. 
The equation 
(\ref{eqn:pntransdiff}) shows that we need only the explicit form of
its  transverse part, say  $\beta^{(3)i}_T$,  for the situation 
in which we are interested here. 
We note that since the gauge 
condition (\ref{eqn:a}) implies that the transverse part
of the  shift vector is a gauge invariant quantity, 
no ambiguities caused by the gauge freedom  remains.   
The perturbation quantity $\beta_T^{(3)i}$ is constrained by
the Einstein equation (\ref{eqn:bm}):
\begin{eqnarray}
 \Delta_x\beta_T^{(3)i}
  =4\left(\frac{ \partial \phi_{N,i}}{\partial t}+
\frac {\dot{a}}{a}\phi_{N,i}\right)+16 \pi G a^2 \rho
   v_N^i.\label{eqn:shifttrans}
\end{eqnarray}
The right-hand side of this equation is shown to be 
divergenceless  through the continuity equation
(\ref{eqn:bt}) at  Newtonian order.
We have already obtained an
expression for the peculiar velocity field and the peculiar gravitational
potential (\ref{eqn:map}) and (\ref{eqn:newtonpot}) in terms of 
the Lagrangian coordinate, respectively.
Therefore, it is adequate to make use of Eq.
(\ref{eqn:shifttrans}) for our purpose. 
Then if we  express Eq. (\ref{eqn:shifttrans}) in
terms of the independent variables $\bm{X}$ and $t$, we can obtain 
the following equations in  simple form:
\begin{eqnarray}
&&\hspace{-1em}\Delta_X\beta^{(3)i}_T-\lambda
  q_z\Psi^{(1)}_{N|jjk}\beta^{(3)i}_T{}_{|k}+3\lambda
  q_z\Psi^{(1)}_{N|jj}\beta^{(3)i}_T{}_{|kk}-2\lambda
  q_z\Psi^{(1)}_{N|jk}\beta^{(3)i}_T{}_{|jk}\nonumber \\
&&\hspace{0em}=\lambda^2 
\frac{4}{t_I^3(1+z_I)^2}\left(\frac{t}{t_I}\right)^{-1/3}\!\left(
-\Psi^{(1)}_{N|jj}\Psi^{(1)}_{N|i}+\Psi^{(1)}_{N|ij}\Psi^{(1)}_{N|j}
+\Psi^{(2)}_{N|i}\right)+O\!\left(\lambda^3\right),\label{eqn:shiftdiffer}
\end{eqnarray}
where we have used Eqs. (\ref{eqn:newtrajsol}), (\ref{eqn:newtonpot})
 and the 
equation of mass conservation, (\ref{eqn:db}). 
We are now in a position to be able to solve for 
the shift vector  iteratively. 
Accordingly, we can conclude that the shift vector $\beta^{(3)i}_T$ is
of  second order in $\lambda$ and obtain
the following equation at  order $\lambda^2$ 
of Eq.(\ref{eqn:shiftdiffer}): 
\begin{eqnarray}
\Delta_X\beta^{(3)i}_T=\lambda^2 
\frac{4}{t_I^3(1+z_I)^2}\left(\frac{t}{t_I}\right)^{-1/3}
\left(-\Psi^{(1)}_{N|jj}\Psi^{(1)}_{N|i}+\Psi^{(1)}_{N|ij}\Psi^{(1)}_{N|j}
+\Psi^{(2)}_{N|i}\right).\label{eqn:shiftpoisson}
\end{eqnarray}
The consistency of our  formulation can easily be checked 
if one confirms  that  the right-hand side of this equation also
 satisfies     
the  divergenceless condition with respect to
the Lagrangian coordinate  by using Eq. (\ref{eqn:newvelsecond}). 
If we again use the Green function of the Laplacian, we can express
the solution of this equation as
\begin{equation}
 \beta^{(3)i}_T(\bm{X},t)= \lambda^2\frac{4}{t_I^3(1+z_I)^2}
  \left(\frac{t}{t_I}\right)^{-1/3}\bar{\beta}_T^{(3)i}(\bm{X})
  +O\!\left(\lambda^3\right) \label{eqn:shiftsol},
 \end{equation}
where
\begin{eqnarray}
\bar{\beta}_T^{(3)i}(\bm{X}):= \int\! d^3\!\bm{Y}\frac{
\left(\Psi^{(1)}_{N|jj}(\bm{Y})\Psi^{(1)}_{N|i}(\bm{Y})
  -\Psi^{(1)}_{N|ij}(\bm{Y})\Psi^{(1)}_{N|j}(\bm{Y})
-\Psi^{(2)}_{N|i}(\bm{Y})\right)}{4\pi |\bm{X}-\bm{Y}|}.\label{eqn:shifspatial}
\end{eqnarray}
%
In the above expression of the integration, we have again assumed that 
the homogeneous solution of Eq. (\ref{eqn:shiftpoisson}) is 
zero.
It should be noted that 
we consider $\beta^{(3)i}$, not $\beta^{(3)}_i$, 
throughout this paper. If one needs the quantity $\beta^{(3)}_i$, 
one finds that it  has a growing mode by the definition 
$\beta^{(3)}_i=a^2\beta^{(3)i}$. 
The above equation  is the expression of the shift vector 
that we have sought in order to solve Eq. (\ref{eqn:pntransdiff}). 
Thus we are able to express all perturbation quantities of the metric 
in our coordinates  in terms of the trajectory field. 
This must be the case  because it is only a dynamical variable 
in the Lagrangian description.  

By inserting Eq. (\ref{eqn:shiftsol}) into Eq. 
(\ref{eqn:pntransdiff}),
we can obtain the following
equation at the lowest order of $\lambda$:
\begin{equation}
 \left(\ddot{Q}^T_z+2\frac{\dot{a}}{a}\dot{Q}^T_z\right)\nabla_X\times
  {}^{(PN)}\!\bm{\Xi}^{(1)}(\bm{X})=\mbox{{\bf 0}}.
\end{equation}
As an irrotational  case, the form of this equation  allows us to seek 
solutions of the form
\begin{equation}
 \ddot{Q}^T_z+2\frac{\dot{a}}{a}\dot{Q}^T_z=0, \label{eqn:pntransfirst}
\hspace{1em}
\mbox{with}\hspace{1em}
 {}^{(PN)}\!\bm{\Xi}^{(1)}(\bm{X})=\bm{\Pi}(\bm{X}),\hspace{2em}\nabla_X\cdot\bm{\Pi}=0,
\end{equation}
where $\bm{\Pi}$ is an unknown function determined by the initial
conditions. The solutions of Eq. (\ref{eqn:pntransfirst}) 
can be easily found to have  only a  decaying mode. The first
order solution  may be safely
 ignored, because the decaying mode plays no  physically 
 important role:
\begin{equation}
 Q^T_z{}^{(PN)}\Xi^{(1)}_i\approx 0.\label{eqn:pntransfirstsol}
\end{equation}
Similarly, by inserting  Eqs. (\ref{eqn:shiftsol}) and
(\ref{eqn:pntransfirstsol}) 
into Eq. (\ref{eqn:pntransdiff}),
we can obtain the following equation at
 second order in $\lambda$:
\begin{equation}
 \left(\ddot{Q}^T_{zz}+2\frac{\dot{a}}{a}\dot{Q}^T_{zz}\right)
   \nabla_X\times {}^{(PN)}\!\bm{\Xi}^{(2)}=-\frac{4}{t_I^4(1+z_I)^2}
 \left(\frac{t}{t_I}\right)^{-4/3}\nabla_X \times\bar{\bm{\beta}}^{(3)}_T.
\label{eqn:pnvortdiff}
\end{equation}
The solution of this equation can be found by solving
the linear ordinary differential equations
\begin{equation}
 \ddot{Q}^T_{zz}+2\frac{\dot{a}}{a}\dot{Q}^T_{zz}=\frac{4}{t_I^2}
  \left(\frac{t}{t_I}\right)^{-4/3},\label{eqn:transsecond}
\end{equation}
\begin{eqnarray}
&&\hspace{0em}\mbox{with}
\hspace{1em}\Delta_X{}^{(PN)}\!\Xi^{(2)}_i
 =-\frac{1}{(1+z_I)^2t_I^2}\left(-\Psi^{(1)}_{N|jj}\Psi^{(1)}_{N|i}
 +\Psi^{(1)}_{N|ij}\Psi^{(1)}_{N|j}+\Psi^{(2)}_{N|i}\right)\label{eqn:pntranspot}.
\end{eqnarray}
Since we expect the quantity $Q^T_{zz}{}^{(PN)}\!\Xi^{(2)}_i$ to be much 
smaller than the order $\lambda$ quantities such as $\mal{\delta}_g$ at
the initial time,  
we can safely adopt  the additional  initial condition 
$\dot{Q}^T_{zz}(t_I){}^{(PN)}\!\Xi^{(2)}_i=0$ as a longitudinal
perturbation case \cite{tf} besides the initial condition $Q^T_{zz}(t_I)=0$.
Then we can find the solution of Eq. (\ref{eqn:transsecond}),
\begin{eqnarray}
 Q^T_{zz}=\left(\frac{3}{2}\right)\left[4\left(\frac{t}{t_I}\right)^{2/3}-12
 +8\left(\frac{t}{t_I}\right)^{-1/3}\right].\label{eqn:pntranstime}
\end{eqnarray}

\section{Discussion and summary}

We have found that 
 the transverse part of the first PN displacement vector has a growing
mode
even if the Newtonian trajectory field does not have a transverse mode. 
We emphasize that the PN solution with the growing mode is a  particular
solution of the differential equation (\ref{eqn:pnvortdiff}) 
and is generated by the initial density fluctuation field through
Eq. (\ref{eqn:newvelpot}).    
Actually, we have found that the longitudinal part of the PN
displacement vector first appears of the  same order of 
its magnitude.\cite{tf}  
Thus the existence of the growing transverse part in the PN displacement
vector  is interesting because 
such a mode is present in neither  Newtonian theory  
nor the   gauge invariant linear theory.

Buchert has shown that the Newtonian trajectory field has 
 no growing transverse mode  
up to  third order in $\lambda$ 
when the initial velocity field is assumed to be
irrotational.\cite{buchert94}        
In his work,
 the solution of the transverse part $q^T_{zzz}{}^{(N)}\!\bm{\Xi}^{(3)}$  
at  order $\lambda^3$ takes  the
 form
\begin{subequations}\label{eqn:newthirdvort}
\begin{eqnarray}
&&\hspace{0em}q^T_{zzz}=\!\left(\frac{3}{2}\right)^3\!\left[\frac{1}{14}
\left(\frac{t}{t_I}\right)^{2}
\!\!-\frac{3}{14}\left(\frac{t}{t_I}\right)^{4/3}\!\!
+\frac{1}{10}\left(\frac{t}{t_I}\right)^{2/3}
\!\!+\frac{1}{2}
-\frac{4}{7}\left(\frac{t}{t_I}\right)^{-1/3}\!\!
+\frac{4}{35}\left(\frac{t}{t_I}\right)^{-1}\right],\nonumber \\ 
\end{eqnarray}
with
\begin{equation}
\Delta_X{}^{(N)}\!\Xi^{(3)}_{i}=\Psi^{(1)}_{N|ijk}\Psi^{(2)}_{N|jk}-\Psi^{(1)}_{N|kjj} 
 \Psi^{(2)}_{N|ki}+\Psi^{(1)}_{N|ki}\Psi^{(2)}_{N|kjj}
-\Psi^{(1)}_{N|kj}\Psi^{(2)}_{N|kij}.  
\end{equation}
\end{subequations}
%
He has  pointed out that the existence of the above transverse part ensures 
the conservation of the vorticity flow along the fluid flow 
up to  order $\lambda^3$.

To see the  importance of the PN effect on  Newtonian dynamics, 
let us make a simple order estimation. Consider 
the initial density fluctuation $\mal{\delta}_{g(l)}$ with the
characteristic length $l$ in the comoving frame. Note that 
due to $a(t_0)$, we may regard the length $l$ as the physical scale at
the present time $t_0$. By using Eqs. (\ref{eqn:newvelpot}),
(\ref{eqn:newvelsecond}), 
(\ref{eqn:pntranspot}), (\ref{eqn:pntranstime}),and (\ref{eqn:newthirdvort}), the above 
 transverse parts of the Newtonian displacement vector 
and the PN displacement vector 
 may be estimated respectively roughly at some
time as 
\begin{eqnarray}
&&q^T_{zzz}{}^{(N)}\!\Xi^{(3)}_{(l)i}\sim
 a^3\left(\delta_{g(l)}\!(t_0)\right)^3,\nonumber \\
&&\frac{1}{c^2}Q^T_{zz}{}^{(PN)}\!\Xi^{(2)}_{(l)i}\sim
a\left(\frac{l}{ct_0}\right)^2\left(\delta_{g(l)}\!(t_0)\right)^2, \label{eqn:orderest}
\end{eqnarray}
where $\delta_{g(l)}\!(t_0)$ is defined by $(1+z_I)\mal{\delta}_{g(l)}$
representing the present value of the density contrast for the scale $l$ 
extrapolated by the linear theory.  
This expression (\ref{eqn:orderest}) suggests the following
interpretations. As long as we consider the evolution of the 
transverse part with scale $l$ much smaller than the present horizon 
scale $ct_0$, it is described well only by Newtonian theory, including 
the transverse part, because 
the Newtonian transverse part $q_{zzz}^T{}^{(N)}\!\Xi^{(3)}_{i}$
remains  much 
larger than the PN transverse part  $Q^T_{zz}{}^{(PN)}\!\Xi^{(2)}_i$ 
from the initial time. This may not be correct for fluctuations with 
 larger scale. In fact, from Eq. (\ref{eqn:orderest}) 
we can evaluate the redshift $z_N$ when the
magnitude of the Newtonian transverse part begins to become larger than
that of the PN transverse part 
 for the density fluctuation field with scale $l$:
\begin{equation}
1+z_N\sim\frac{ct_0}{l}\left(\delta_{g(l)}\!(t_0)\right)^{1/2}. 
\end{equation}
Noting that $ct_0$ is on the order of  the present horizon scale and 
$\sim2000h^{-1}\mbox{Mpc}$, 
it is found that 
this redshift is smaller for  fluctuations with larger scale.  
This means that the PN transverse part might play an important role for
the early evolution of large-scale structure. 
Since  some fluctuations with large scale  
become localized smaller in comparison  with the horizon scale, 
which grows larger, 
it is naturally described more accurately by Newtonian dynamics later.  
For example, SDSS is expected to survey a cubic region of several hundred
megaparsecs length. Then the above estimate gives $z_N\sim 4$ for
fluctuations with  scale $50h^{-1}\mbox{Mpc}$ at the present time, 
if we estimate the magnitude of the present density contrast  as 
$\delta_{g(50)}\!\!(t_0)\sim 10^{-2}$, according to 
 the power spectrum of the density contrast  
in the standard CDM model\cite{bardeen86} normalized by the 
COBE observation.  

We can solve for the trajectory field, namely  
the integral curve of velocity field $\bm{v}$ in our coordinates, up 
to the 1st PN order, so it will be interesting to see how 
the relativistic effect on Newtonian dynamics appears as a
characteristic pattern in the large-scale structure formation. 
It will be easy to investigate this problem expanding 
 the Newtonian simulation based on the
Lagrangian perturbation approach.     
Also, as stated above, the PN characteristic effect might 
appear on larger scales. 
In particular, the effect may induce  secondary temperature anisotropies 
of CMB,  as the Rees-Sciama effect,\cite{rees} and potentially be observable.  
This reason in the anisotropy may be  interpreted as follows. 
As stated in \S 2, we foliate the spacetime 
by  simultaneous surfaces of the basic observers who see no dipole
component  of 
CMB using  the (3+1) formalism. 
Thus, it would  be natural  for 
the existence  of the shift vector to influence the anisotropy of CMB.  
These  works are  now in progress. 

Since the observable region of the large-scale structure increases
steadily, we expect that our work will play some important 
role in the future.  


\end{document}